# High-Pressure Synthesis of Barium Superhydrides: Pseudocubic BaH$_{12}$

Wuhao Chen,[1] Dmitrii V. Semenok,[2] Alexander G. Kvashnin,[2,*] Ivan A. Kruglov,[3,4] Michele Galasso,[2] Hao Song,[1] Xiaoli Huang,[1,*] Defang Duan,[1] Alexander F. Goncharov,[5] Vitali B. Prakapenka,[6] Artem R. Oganov,[2,7,*] and Tian Cui[8,1,*]

[1] State Key Laboratory of Superhard Materials, College of Physics, Jilin University, Changchun 130012, China

[2] Skolkovo Institute of Science and Technology, Skolkovo Innovation Center, 3 Nobel Street, Moscow 143026, Russia

[3] Moscow Institute of Physics and Technology, 9 Institutsky Lane, Dolgoprudny 141700, Russia

[4] Dukhov Research Institute of Automatics (VNIIA), Moscow 127055, Russia

[5] Earth and Planets Laboratory, Carnegie Institution of Washington, 5251 Broad Branch Road NW, Washington, D.C. 20015, U.S.

[6] Center for Advanced Radiation Sources, The University of Chicago, 5640 South Ellis Avenue, Chicago, Illinois 60637, U.S.

[7] International Center for Materials Discovery, Northwestern Polytechnical University, Xi'an 710072, China

[8] School of Physical Science and Technology, Ningbo University, Ningbo, 315211, China

**Corresponding authors**
*Dr. X. Huang, huangxiaoli@jlu.edu.cn
*Prof. T. Cui, cuitian@jlu.edu.cn
*Prof. A.R. Oganov, a.oganov@skoltech.ru

**Abstract**
Following the discovery of high-temperature superconductivity in the La–H system, where for the recently discovered *fcc*-LaH$_{10}$ a record critical temperature $T_C$ = 250 K was achieved [Drozdov et al., *Nature*, 569, 528 (2019); Somayazulu et al., *Phys. Rev. Lett.* 122, 027001 (2019)], we studied the formation of new chemical compounds in the barium-hydrogen system at pressures of 75 to 173 GPa. Using *in situ* generation of hydrogen from NH$_3$BH$_3$, we synthesized previously unknown superhydride BaH$_{12}$ with a pseudocubic (*fcc*) Ba sublattice, which was observed in a wide range of pressures from ~75 to 173 GPa in four independent experiments. DFT calculations indicate a close agreement between the theoretical and experimental equations of state. In addition to BaH$_{12}$, we identified previously known *P*6/*mmm*-BaH$_2$ and possibly BaH$_{10}$ and BaH$_6$ as impurities in the samples. *Ab initio* calculations show that newly discovered semimetallic BaH$_{12}$ contains H$_2$, H$_3^-$ molecular units and detached H$_{12}$ chains. Barium dodecahydride is a unique molecular hydride with metallic conductivity which demonstrates a superconducting transition around 20 K at 140 GPa in agreement with calculations (19–32 K). The interpretation of the multiphase XRD data was possible thanks to the development of new Python scripts for postprocessing the results of evolutionary searches. These scripts help quickly identify the theoretical structures that explain the experimental data in the best way, among thousands of candidates.

## Introduction

In recent years, the search for new hydride superconductors with $T_C$ close to room temperature, attracts great attention of researchers in the field of high-pressure materials science. Variation of pressure opens perspectives for synthesis of novel functional materials with unexpected properties.[1] For example, according to theoretical models,[2–5] compression of molecular hydrogen over 500 GPa should lead to the formation of an atomic metallic modification with $T_C$ near room temperature. Pressures of 420-480 GPa were achieved in experiments with toroidal diamond anvil cells;[6] however, for conventional high-pressure cells with a four-electrode electric setup, pressures above 200 GPa remain challenging.

In 2004, Ashcroft[7] suggested an alternative method of searching for high-$T_C$ superconductors using other elements, metals or nonmetals, to precompress the hydrogen atoms, which should lead to a dramatic decrease in the metallization pressure. At the same time, because H is the lightest element in the periodic table, these hydrogen-dominated compounds may have a high Debye temperature and strong electron-phonon coupling. Ashcroft's work triggered many related theories and experiments.

A decade later Ashcroft's idea found its experimental proof. Extraordinarily high superconducting transition temperatures were demonstrated in compressed $Fm\bar{3}m$-$H_3S$[8–11] (203 K at 150 GPa), $Fm\bar{3}m$-$LaH_{10}$[12–14] (>250 K at 175 GPa), $Im\bar{3}m$-$YH_6$[15] and $P6_3/mmc$-$YH_9$[16] (224 and 243 K, respectively), $Fm\bar{3}m$-$ThH_{10}$[17] (161 K at 174 GPa), and $P6_3/mmc$-$CeH_9$ (~100 K).[18] Recently, several semiempirical criteria for search for new hydride superconductors were proposed, for example, the clathrate (H-cage) hydrogen substructure[19,20] and "lability belt" of superconductivity.[21] From the theoretical point of view, only several metal polyhydrides satisfy these criteria and have calculated $T_C$ above or close to 200 K. Mostly these are hydrides of alkali earth (Mg, Ca, Sr, Ba, etc.) and early rare earth (La, Y, and Sc) metals. X-ray diffraction (XRD) experiments on compounds with heavy atoms are more convenient because of better signal-to-noise ratio than on those with light elements like Li, Be, and Mg. Another practical reason to study hydrides of heavy metals is that their stabilization requires lower pressure. The minimum stabilization pressure determines the maximum sample size that can be loaded into a DAC, which, together with the atomic number, determines the exposure time and contrast of the diffraction pattern. These reasons make the XRD experiments with polyhydrides of light elements complicated.

The neighbor of lanthanum, barium is a promising element for the superhydride synthesis. The calculated maximum $T_C$ is only about 30–38 K[20,21] for tetragonal $P4/mmm$-$BaH_6$ stable at 100 GPa, which has a hydrogen sublattice consisting of $H_2$ molecules and $H^-$ anions.[20] Lower barium hydride, $BaH_2$, well-known for its extraordinarily anionic ($H^-$) conductivity,[22] exists in $Pnma$ modification below 2.3 GPa, whereas above 2.3 GPa it undergoes a transition to hexagonal $Ni_2In$-type $P6_3/mmc$ phase.[23] Chen et al.[24] proposed a $YbZn_2$-type structure of $BaH_2$ with $Imma$ space group on the basis of ab initio calculations. However, this compound has not yet been observed in experiment. At pressures above 41 GPa, $BaH_2$ transforms to $P6/mmm$ modification, which metallizes at over 44 GPa, but its superconducting $T_C$ is close to zero.[25]

None of the previously proposed structures of stable barium polyhydrides have a clathrate hydrogen sublattice. So far, no relevant experiments at pressures above 50 GPa have been reported. Due to analogy of the Ba–H and La–H systems, a recalculation of the Ba–H system is required and the existence of $BaH_n$, where $n \geq 10$, is expected (for instance, $BaH_{10}$ and $BaH_{12}$). Closing the gap of previous studies, in this work we experimentally and theoretically investigated the chemistry of the barium-hydrogen system at pressures from 75 to 173 GPa.

## Results and Discussion

To investigate the formation of new chemical compounds in the Ba–H system at high pressures, we loaded four high-pressure diamond anvil cells (DACs #B0–B3) with a sublimated ammonia borane $NH_3BH_3$ (AB), used as a source of hydrogen and a pressure transmitting medium. A tungsten foil preindented to a thickness of about 20 μm was used as a gasket. Additional parameters of the high-pressure diamond anvil cells are given in Table 1.

It has been shown in the synthesis of superhydrides of lanthanum,[13,14] thorium,[17] praseodymium,[26] and neodymium[27] that ammonia borane is an effective source of hydrogen that releases it when the metal target is heated by a short (< 0.1 s) laser pulse. The generation of hydrogen occurs due to the well-known decomposition reaction: $NH_3BH_3 \rightarrow 3H_2 + c\text{-}BN$.[28]

Table 1. Experimental parameters of the DACs that were used to synthesize barium hydrides

| Cell # | Synthesis pressure, GPa | Culet size, μm | Sample size, μm | Composition/load |
|---|---|---|---|---|
| B0 | 173 | 50 | 17 | Ba/BH$_3$NH$_3$ |
| B1 | 160 | 50 | 18 | Ba/BH$_3$NH$_3$ |
| B2 | 146 | 100 | 25 | Ba/BH$_3$NH$_3$ |
| B3 | 90 | 100 | 20 | Ba/BH$_3$NH$_3$ |

# 1. Synthesis at 160 GPa and stability of BaH$_{12}$

The first attempt of the experimental synthesis was performed in DAC #B1 heated to 1700 K by an infrared laser pulse of ~0.5 s at a pressure of 160 GPa. During heating the Ba particle undergoes a significant expansion and remains nontransparent (Figure 1b,c). The obtained synchrotron X-ray diffraction pattern (λ = 0.62 Å, Figure 1a) consists of a series of strong reflections specific to cubic crystals. Some additional weak reflections were also detected, which may point to a distortion of cubic barium sublattice. Decreasing the pressure in the DAC #B1 to 119 GPa (Figure 3a) gives a series of diffraction patterns which can mostly be indexed by a slightly distorted cubic structure (e.g., $Cmc2_1$, as shown in Figure 1a; hexagonal space groups cannot be used for the refinement). Recently, similar diffraction patterns have been observed above 150 GPa for the La–H ($fcc$-LaH$_{10}$[13,14]) and Th-H ($fcc$-ThH$_{10}$[17]) systems. By the analogy with the La–H system, and considering the lack of previously predicted cubic superhydrides BaH$_x$,[20,22,21] we performed theoretical crystal structure evolutionary search for stable Ba–H compounds using the USPEX code[29–31] at pressures of 100–200 GPa and temperatures of 0–2000 K, both variable- and fixed-composition searches.

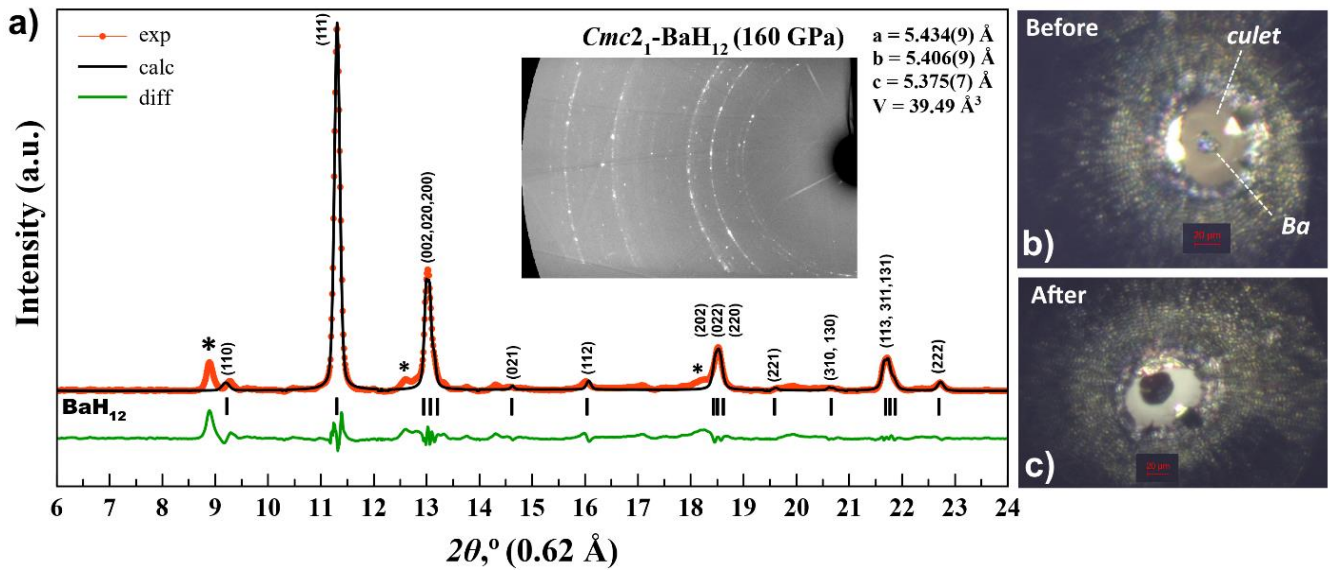

**Figure 1.** Experimental X-ray diffraction pattern from DAC #B1 at 160 GPa and the Le Bail refinement of the pseudocubic $Cmc2_1$-BaH$_{12}$ phase (a). The experimental data, fitted line, and residues are shown in red, black, and green, respectively. Unidentified reflections are indicated by asterisks. Microphotographs of the cell #B1 culet with a Ba sample before (b) and after (c) the laser heating. A substantial increase in the area of the sample and a change in the reflectivity occurred.

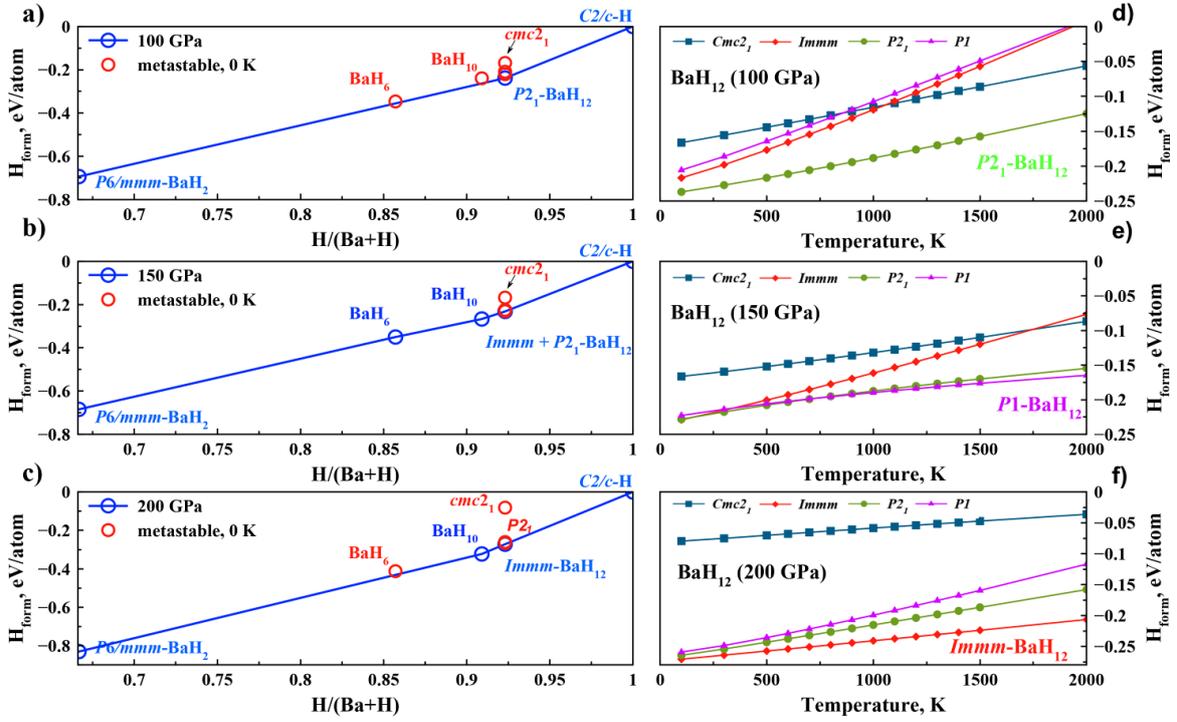

**Figure 2.** Convex hulls of the Ba–H system at (a) 100, (b) 150, and (c) 200 GPa calculated with zero-point energy (ZPE) contribution. Dependence of the Gibbs free energies of formation on the temperature for different modifications of $BaH_{12}$ — $Immm$, pseudocubic $Cmc2_1$, $P2_1$, and $P1$ — calculated within the harmonic approach in the temperature range of 100 to 2000 K at (d) 100, (e) 150, and (f) 200 GPa.

According to the USPEX calculations, $P6/mmm$-$BaH_2$ remains stable up to 150–200 GPa (Figure 2; Supporting Information Tables S4–S9, Figures S1 and S2). $BaH_2$ was experimentally detected in DAC #B0 at 173 and 154 GPa with the cell volume by 5% smaller than theoretically predicted (Supporting Information Table S11).

At 100–200 GPa, several new barium polyhydrides lying on or near the convex hulls were found: $BaH_6$, $BaH_{10}$, and $BaH_{12}$ (the unit cell is $Ba_4H_{48}$ or $Ba_8H_{96}$ (Figure 2)). In the experiments at 142 and 154–173 GPa we detected a series of reflections that can be indexed by $BaH_6$ and $BaH_{10}$ with the unit cell volumes close to the calculated ones (see Supporting Information). However, the main phase in almost all diffraction patterns is pseudocubic $BaH_{12}$.

The analysis of the experimental data within $Fm\bar{3}m$ space group (Table 2, Figure 3), compared to DFT calculations, shows that the stoichiometry of barium hydride synthesized in DAC #B1 is close to $BaH_{12}$. Examining the results of the fixed-composition search, we found that pseudocubic $P2_1$-$BaH_{12}$, whose predicted diffraction pattern is similar to the experimental one, lies on the convex hull at 100 GPa. There are also pseudocubic $P1$-$Ba_8H_{96}$, located close to the convex hull at 150 GPa, and $Cmc2_1$-$BaH_{12}$ (or $Ba_4H_{48}$), with the a similar XRD pattern, lying a bit further from the convex hull. Other possible candidate, orthorhombic $Immm$-$BaH_{12}$, which stabilizes between 150 and 200 GPa, does not correspond to the experimental XRD (Figure 1) and was not further considered.

**Table 2.** Experimental cell parameters of the refined pseudocubic $Cmc2_1$-$BaH_{12}$ structure (Figures 1 and 3). Volumes are given in Å³ per Ba atom. When the pressure decreased below 119 GPa, the diamond anvil cell was broken.

| Pressure, GPa | $a$, Å | $b$, Å | $c$, Å | $V$, Å³ | $V_{DFT}$, Å³ |
|---|---|---|---|---|---|
| 160 | 5.434(9) | 5.406(9) | 5.375(7) | 39.49 | 39.48 |
| 154 | 5.442(5) | 5.417(3) | 5.384(1) | 39.69 | 39.99 |
| 145 | 5.480(8) | 5.416(3) | 5.453(2) | 40.47 | 40.80 |
| 135 | 5.485(5) | 5.502(6) | 5.463(3) | 41.22 | 41.77 |
| 126 | 5.561(0) | 5.507(1) | 5.475(5) | 41.92 | 42.74 |
| 119 | 5.585(9) | 5.537(9) | 5.520(3) | 42.69 | 43.53 |

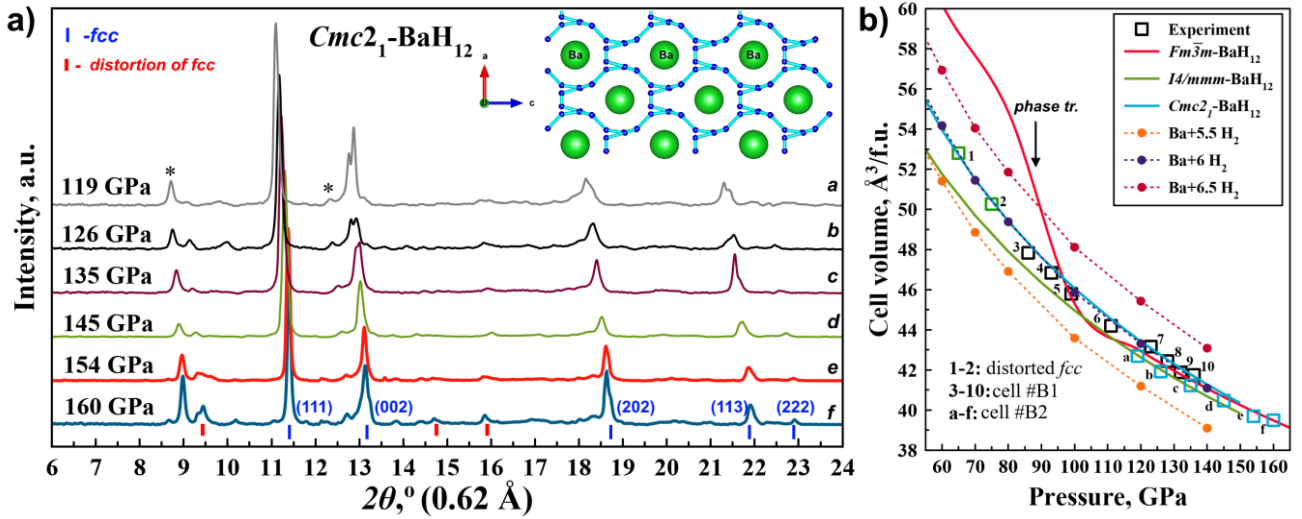

**Figure 3.** Experimental XRD patterns from cell #B1 at pressures of 119 to 160 GPa (a). Unidentified reflections are indicated by asterisks. Inset shows the projection of the structure to the (ac) plane. Cyan thick lines show the hydrogen network. (b) Equations of state for different possible crystal modifications of $BaH_{12}$ (*fcc*, *I4/mmm*, and *Cmc*$2_1$) and Ba+$n$H$_2$.

Several cubic structures may be proposed to explain the experimental XRD patterns, for instance, $Fm\bar{3}m$-$BaH_{12}$ (similar to *fcc*-$YB_{12}$) or $Pm\bar{3}$-$BaH_{12}$. However, the latter one has a much lower theoretical (DFT) volume compared to the experimental one (e.g., at 160 GPa the calculated volume of $Pm\bar{3}$-$BaH_{12}$ is 37.68 Å$^3$, whereas the experimental one is 39.49 Å$^3$). The computed equation of state of $Fm\bar{3}m$-$BaH_{12}$ (Figure 3b) corresponds much better to the experimental volume-pressure dependence. However, the DFT calculations show that the ideal $Fm\bar{3}m$ barium sublattice is unstable both thermodynamically and dynamically, and transforms spontaneously to $Cmc2_1$ or $P2_1$ via distortion (Figure 4). Moreover, high-symmetry cubic phases cannot explain the presence of weak reflections at 8.9–9.4°, 14.5, 16, 19.5, and 20.6° that can be seen in many XRD patterns (Figure 1 and Figure 3, and Supporting Information).

Tetragonal $I4/mmm$-$BaH_{12}$ (unit cell is $Ba_2H_{24}$, its structure is presented in Supporting Information Table S1) may explain most of the weak reflections, just like distorted pseudocubic (where $a \approx b \approx c$) $P2_1$-$BaH_{12}$ and $Cmc2_1$-$BaH_{12}$, but $P2_1$-$BaH_{12}$ lies much closer to the convex hull at 100–150 GPa and is close to achieving the dynamic stability (in the harmonic approximation, see below). Studying the dependence of the Gibbs free energy of formation on the temperature (Figure 2d–f), we found that $P2_1$-$BaH_{12}$ is the most stable modification at 0–2000 K and 100–150 GPa. At the same time, tetragonal $BaH_{12}$, lying more than 0.1 eV/atom above the convex hull, is unstable both dynamically and thermodynamically.

To clarify the question of dynamic stability of pseudocubic structures, we calculated a series of phonon densities of states for different modifications of $BaH_{12}$ (Figure 4). Symmetrical and corresponding to the experimental data $Cmc2_1$-$BaH_{12}$ has a number of imaginary phonon modes (within the harmonic approach). Its distortion to thermodynamically stable $P2_1$-$BaH_{12}$ leads to the disappearance of many of the imaginary phonon modes, an increase in the deviations with the experimental XRD pattern (peak splitting), and an emergence of a pseudogap in the electronic density of states $N(E)$ (Figure 4e). A subsequent distortion of $P2_1$ to $P1$ converts $BaH_{12}$ to a semiconductor with a bandgap over 0.5 eV (the systematic underestimation of a bandgap by the DFT should be taken into account). However, the experimental data show that $BaH_{12}$ remains an opaque (in the visible range) material with an almost *fcc* crystal structure and exhibits metallic properties (see next sections) down to 75 GPa. For this reason, the band structure and parameters of the superconducting state were further investigated only for $Cmc2_1$-$BaH_{12}$, which does not have a bandgap at 100–150 GPa.

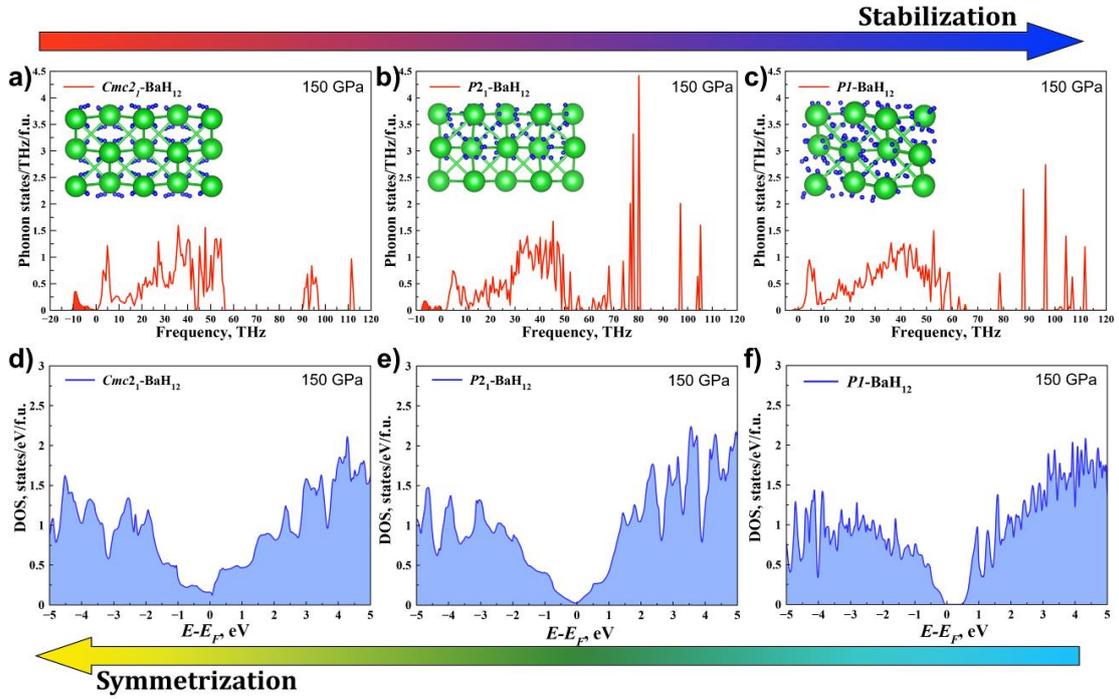

**Figure 4**. Phonon and electron densities of states for pseudocubic $BaH_{12}$ structures with various degrees of distortion: (a, d) $Cmc2_1$, (b, e) $P2_1$, and (c, f) $P1$.

Comparative analysis of $Cmc2_1$, $P2_1$, and $P1$ structures of $BaH_{12}$ shows that semimetallic $Cmc2_1$ best explains the experimental results of the X-ray diffraction, and lies much closer to the convex hull than $Fm\bar{3}m$ or $I4/mmm$ modifications. $P2_1$-$BaH_{12}$ shows a complex picture of splitting the diffraction signals, has a bandgap of ~0.5 eV at 100 GPa, and its equation of state contradicts the experimental data. Therefore, pseudocubic $Cmc2_1$-$BaH_{12}$, whose cell volume is near that of the close-packed $Fm\bar{3}m$-$BaH_{12}$, is the appropriate explanation of the experimental results despite its instability. More accurate analysis accounting for the anharmonic nature of hydrogen oscillations, as it was done for $fcc$-$LaH_{10}$,[32,33] may explain the experimental stability of higher-symmetry $Cmc2_1$-$BaH_{12}$ compared to lower-symmetry $P2_1$-$BaH_{12}$.

The molecular dynamics simulation of $Cmc2_1$-$BaH_{12}$ and $P2_1$-$BaH_{12}$ at 10–1500 K, after averaging the coordinates, leads to distorted pseudocubic $P1$-$BaH_{12}$ with the same XRD pattern. All structures retrieved by molecular dynamics are less stable both dynamically and thermodynamically than $P1$-$BaH_{12}$, $P2_1$-$BaH_{12}$, and $Cmc2_1$-$BaH_{12}$ found by USPEX.

## 2. Synthesis of $BaH_{12}$ at 146 GPa

Similar X-ray diffraction patterns were obtained in the next experiment (DAC #B2) where Ba sample was heated at an initial pressure of 146 GPa, which led to a decrease in pressure to 140 GPa. During the heating and subsequent unloading of the cell, the sample remained opaque down to ~40 GPa. Unlike the synthesis at high pressure (cell #B1, 160 GPa), in this experiment we observed much more side phases and corresponding side reflections than before (Figure 5 and Supporting Information). This corresponds with the fact that higher pressures make it possible to synthesize hydrides with a higher hydrogen content, whereas lower pressures lead to the dominance of lower hydrides in the reaction mixture.[13, 15-17]

Similar to the experiment with DAC #B1, five reflections from the pseudocubic Ba sublattice dominate over a wide range of pressures (65–140 GPa), whereas side reflections change their intensities and, at some pressures, almost disappear (Figure 6a and Supporting Information Figures S30 and 32). At 65 GPa, the (202)-reflection of the $fcc$-structure almost vanishes from the diffraction pattern, whereas the intensity of a side reflection at 11–11.5°, which is placed between (111) and (002), continues to grow (Supporting Information Figure S30). At pressures below 65 GPa, it is no longer possible to refine the cell parameters of pseudocubic $BaH_{12}$. The parameters of the $Cmc2_1$-$BaH_{12}$ unit cell, refined to the experimental data, are presented in Table 3.

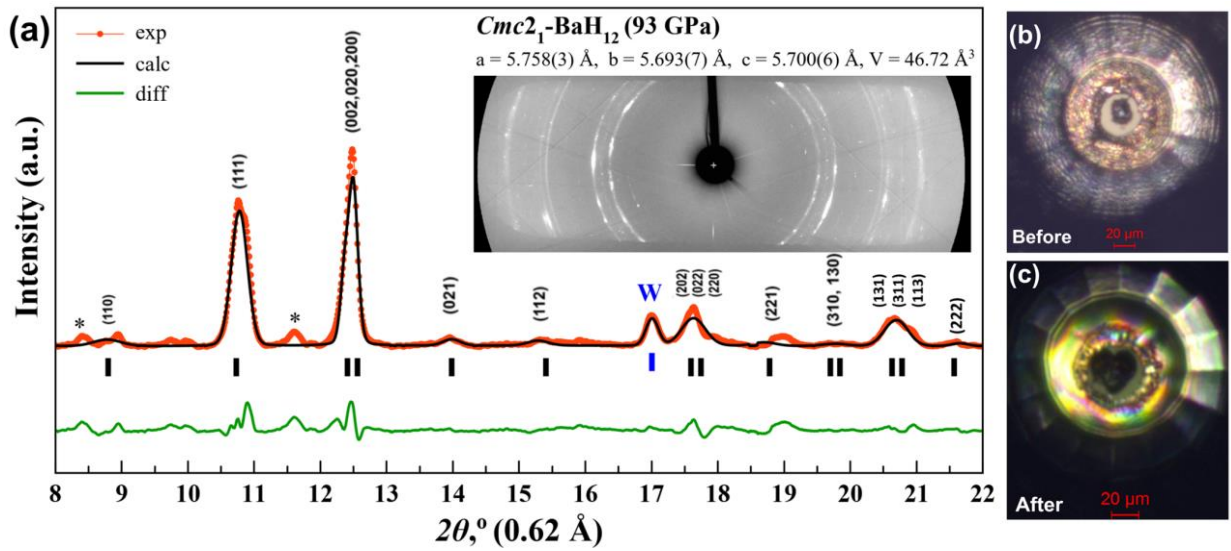

**Figure 5**. Le Bail refinement of pseudocubic $Cmc2_1$-BaH$_{12}$ and $bcc$-W at 93 GPa (a). The experimental data, fitted line, and residues are shown in red, black, and green, respectively. Inset shows the 2D diffraction image. Weak reflections, indicated by asterisks, may correspond to impurity of hexagonal BaH$_{12}$. Microphotographs of the cell #B2 culet with a Ba/AB sample before (b) and after (c) the laser heating.

The diffraction circles corresponding to the ideal cubic barium sublattice have a pronounced granularity, whereas for impurity reflections it is weak for all studied pressure points (Figure 6b–d), which also suggests that all "cubic" reflections belong to the same phase. The impurity peaks between (111) and (020) reflections of pseudocubic BaH$_{12}$ (Figure 6a) may be indexed in $P6_3/mmc$ or $P6_3mc$ space groups, whose cell parameters are listed in Table 3. We also can not exclude formation of complex compounds as result of decomposition of ammonia borane, Ba-AB salts and reactions of N, B with sample that may explain observation of extra reflections in XRD.

Some deviations of the experimental cell volume (~ 1-2%) of $Cmc2_1$-BaH$_{12}$ obtained in different DACs (#B1 and #B2) from the calculated values at 132–138 GPa may be due to either errors or irregularities in pressure measurements when it decreases in small steps, or small variations in stoichiometry (±0.25H). Fitting the experimental pressure-volume data for BaH$_{12}$ in the pressure range from 75 to 173 GPa by the third-order Birch–Murnaghan equation of state[34] gives $V_{100} = 45.47 \pm 0.13$ Å$^3$, $K_{100} = 305 \pm 8.5$ GPa, and $K_{100}' = 3.8 \pm 0.48$. Fitting the theoretical data yields similar values: $V_{100} = 46.0$ Å$^3$, $K_{100} = 315.9$ GPa, and $K_{100}' = 2.94$.

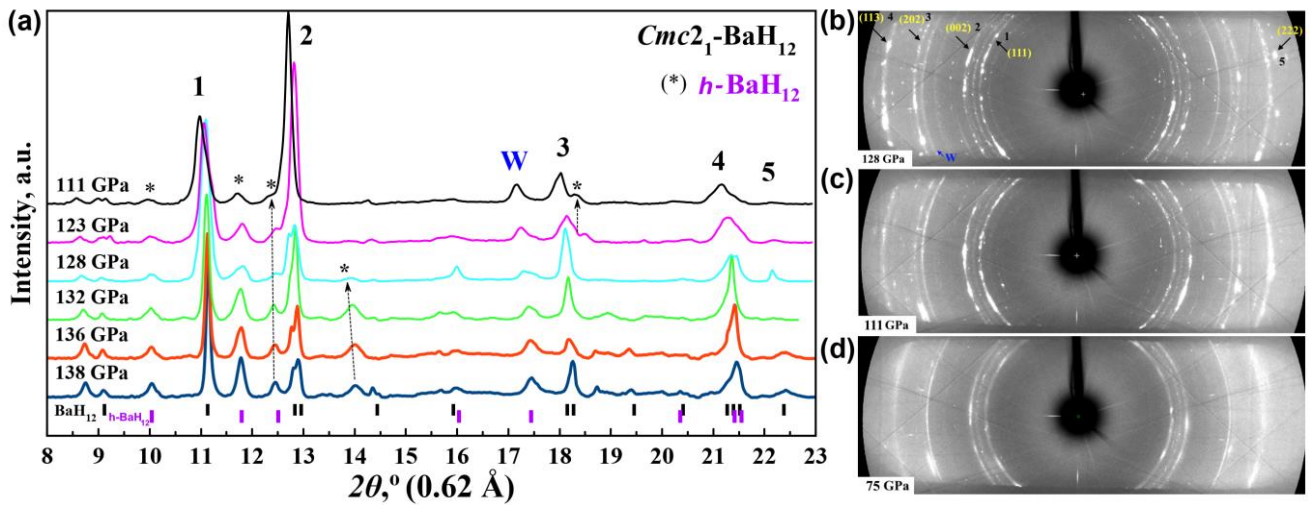

**Figure 6.** Experimental XRD patterns from cell #B2 at pressures decreasing from 140 to 111 GPa (a). Weak reflections from side phases (which can be indexed by $P6_3/mmc$ space group) and decomposition products are indicated by asterisks and dotted lines. (b, c, d) Powder X-ray diffractograms at 128, 111, and 75 GPa, respectively. Reflections corresponding to the crystallographic planes in the ideal $Fm\bar{3}m$-BaH$_{12}$ phase are indicated by arrows.

**Table 3.** Experimental cell parameters of refined pseudocubic $Cmc2_1$-BaH$_{12}$ and proposed $h$-BaH$_{12}$ (impurity). Volumes are given in Å$^3$ per Ba atom. As the pressure decreases below 75 GPa, the refinement using this structure is no longer possible. Below 38 GPa, the diamond anvil cell was broken.

| | Pressure, GPa | $a$, Å | $b$, Å | $c$, Å | $V$, Å$^3$ | $V_{DFT}$, Å$^3$ |
|---|---|---|---|---|---|---|
| $Cmc2_1$-BaH$_{12}$ | 140 | 5.500(1) | 5.481(6) | 5.539(3) | 41.75 | 41.27 |
| | 138 | 5.526(9) | 5.567(4) | 5.500(3) | 42.31 | 41.45 |
| | 136 | 5.535(2) | 5.576(2) | 5.517(7) | 42.58 | 41.68 |
| | 132 | 5.546(4) | 5.595(6) | 5.510(8) | 42.75 | 42.07 |
| | 128 | 5.542(1) | 5.590(4) | 5.505(0) | 42.64 | 42.50 |
| | 123 | 5.560(1) | 5.606(3) | 5.525(9) | 43.06 | 43.05 |
| | 111 | 5.602(6) | 5.638(8) | 5.571(2) | 44.00 | 44.50 |
| | 99 | 5.706(5) | 5.665(6) | 5.637(4) | 45.57 | 46.21 |
| | 93 | 5.758(3) | 5.693(7) | 5.700(6) | 46.72 | 47.12 |
| | 86 | 5.759(9) | 5.812(0) | 5.717(9) | 47.85 | 48.30 |
| | 75 | 5.831(0) | 5.891(8) | 5.797(2) | 49.72 | 50.37 |
| $h$-BaH$_{12}$ | 140 | 4.0798 | 4.0798 | 5.7151 | 41.19 | 40.97 |
| | 136 | 4.0998 | 4.0998 | 5.7151 | 41.59 | 41.32 |
| | 132 | 4.1050 | 4.1050 | 5.7351 | 41.85 | 41.71 |
| | 111 | 4.1250 | 4.1250 | 5.7751 | 42.55 | 43.99 |

## 3. Synthesis at 90 GPa and electronic properties of BaH$_{12}$

In the experiment with DAC #B3, we investigated the possibility to synthesize BaH$_{12}$ at pressures below 100 GPa. After the laser heating of Ba/AB at 1600 K, the pressure in the cell decreased from 90 to 84 GPa and the resulting compounds were investigated using the synchrotron X-ray radiation ($\lambda$ = 0.62 Å). The observed diffraction pattern is generally similar to those in the previous experiments with DAC #B1, except the presence of the impurity phase $h$-BaH$_{\sim 12}$, whose reflections may be indexed by hexagonal $P6_3/mmc$ or $P6_3mc$ space groups (lattice parameters $a$ = 3.955(7) Å, $c$ = 7.650(7) Å, $V$ = 51.84 Å$^3$ at 78 GPa) and whose cell volume corresponds to approximately the same H content as in pseudocubic BaH$_{12}$. For the main set of reflections, slightly distorted cubic BaH$_{12}$ is the best solution (Figure 7). The refined cell parameters of BaH$_{12}$ (see Supporting Information, Table S3) are in agreement with the previously obtained results (DACs #B1 and #B2). When the pressure was reduced to 78 GPa, barium dodecahydride began to decompose, and subsequent diffraction patterns (e.g., at 68 GPa, see Supporting Information) show a complex image of broad reflections that confirms the previously found lower experimental bound of BaH$_{12}$ stability (~75 GPa).

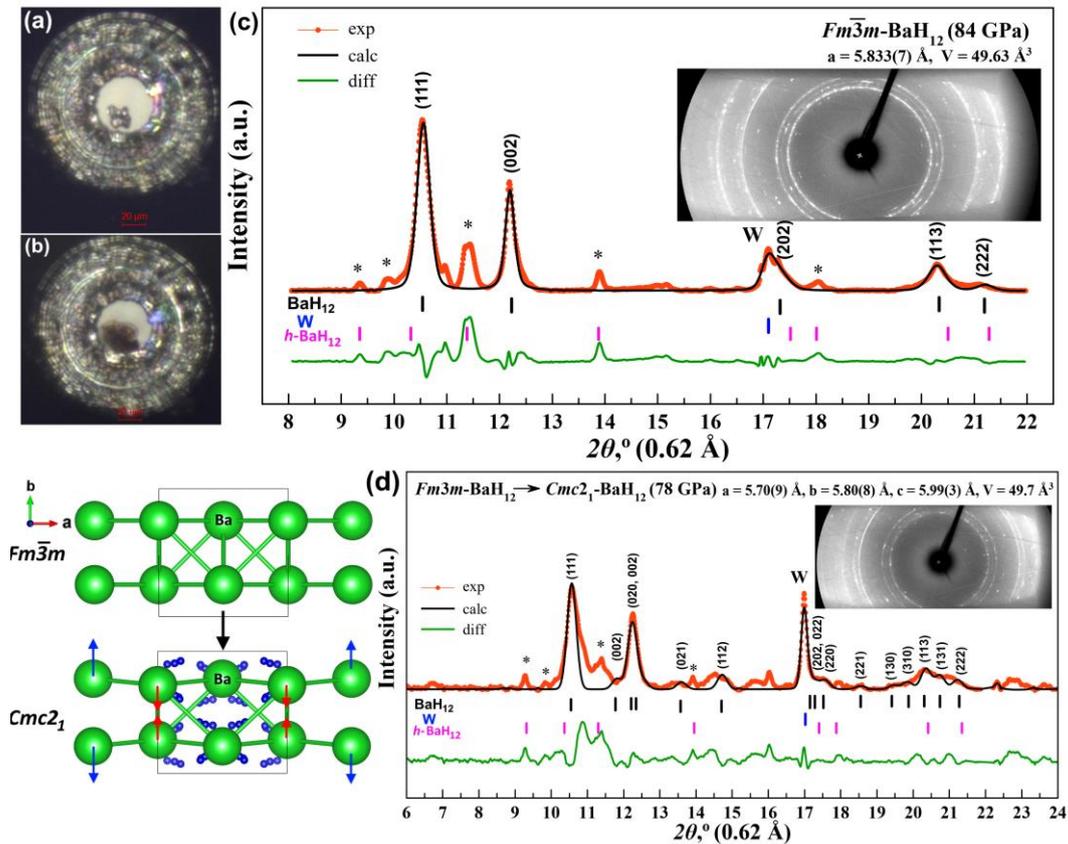

**Figure 7.** Ba/AB sample (a) before and (b) after the laser heating in DAC #B3. Experimental XRD pattern and the Le Bail refinement of (c) the cubic $Fm\bar{3}m$-BaH$_{12}$ structure at 84 GPa and (d) $Cmc2_1$-BaH$_{12}$ structure obtained via the distortion of cubic $Fm\bar{3}m$-BaH$_{12}$ at 78 GPa. The reflections indicated by asterisks may correspond to an unidentified hexagonal barium polyhydride BaH$_{\sim 12}$. The experimental data, model fit for the structure, and residues are shown in red, black, and green, respectively.

BaH$_{12}$ is the first known metal hydride with such a high hydrogen composition that is stable at such low pressures (~75 GPa). We performed further investigation of its electronic structure and the charge state of hydrogen and barium atoms. The electron localization function (ELF) analysis [35] (Figure 8) shows that hydrogen in BaH$_{12}$ divided into groups of H$_2$ ($d_{H-H}$ = 0.78 Å) and H$_3$ ($d_{H-H}$ = 0.81 and 1.07 Å) molecular fragments located close to each other, which are adjacent to barium atoms and form flat horseshoe-like H$_{12}$ chains (all $d_{H-H}$ < 1.27 Å) separated from each other ($d_{12}$ > 1.67 Å, Figure 3). In general, this structure can be represented as a layered crystal with relatively weak links not only between the Ba layers ($d_{Ba-Ba}$ = 2.75 Å), but also between the H layers ($d_{L1-L2}$ = 1.35 Å). If we restrict the consideration of distances up to 1.2 Å, we will see horseshoe-like groups H$_6$ and almost linear H$_3$ groups.

The Bader charge analysis of $Cmc2_1$-BaH$_{12}$, performed in accordance with our previous experience[36,37] (Supporting Information Table S15), shows that the Ba atoms serve as a source of electrons for the hydrogen sublattice. The charge of the barium atoms in BaH$_{12}$ is +1.15 at 150 GPa, whereas most of the hydrogen atoms have a negative charge. In the H$_3$ fragments, the charge of the end atoms is close to –0.2 and –0.27, while the H bridge has a small positive charge of +0.06 (Figure 8a,b). In general, H$_3^-$ anion, similar to one found in the structure of NaH$_7$ [38], has a total charge of –0.4|e|, whereas molecular fragments H$_2$ ($d_{H-H}$ = 0.78 Å) have a charge of only –0.1|e|. Therefore, the Ba–H bonds in BaH$_{12}$ have substantial ionic character, whereas the H–H bonds are mainly covalent.

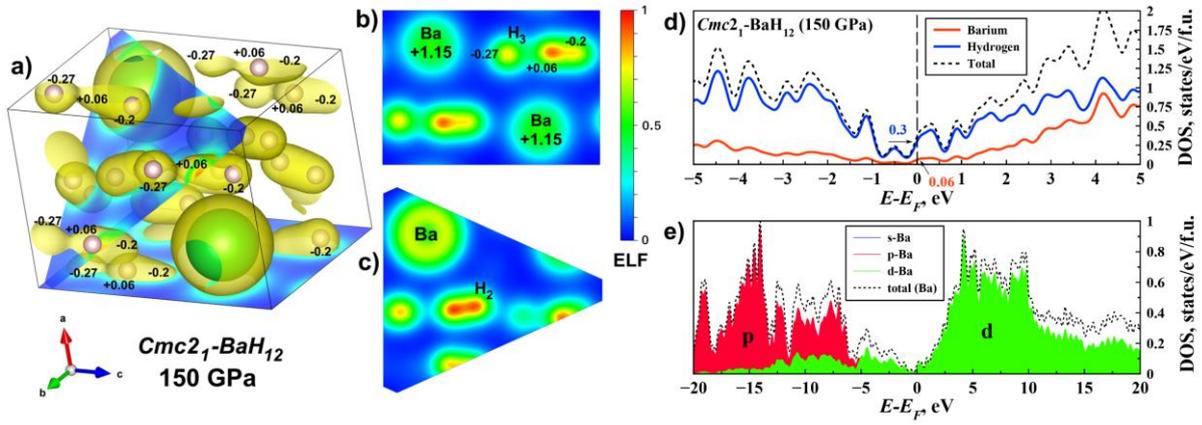

**Figure 8.** (a-c) Electron localization function (ELF is projected onto the (100) plane (b) and the (11-1) plane (c)) and Bader charges of the Ba and H atoms in $Cmc2_1$-$BaH_{12}$ at 150 GPa. (d) Contribution of barium and hydrogen to the electronic density of states of $BaH_{12}$. (e) d-character of Ba electrons near Fermi level.

Spin-polarized calculations demonstrate that in the pressure range of 50–200 GPa all barium hydrides are nonmagnetic. The low electronic density of states $N(E)$ in semimetallic $Cmc2_1$-$BaH_{12}$, which looks typical for one-dimensional …H–H–H… chains (Figure 8d), weakly depends on pressure (Supporting Information Figure S33). The main contribution to $N(E_F)$, 83% at 150 GPa, comes from hydrogen (Figure 8d), and ¾ of this contribution is related to s orbitals. At 150 GPa, barium in $BaH_{12}$ exhibits the properties of d element and its bonding orbitals have a significant d-character (Figure 8e). In the electronic band structure (Supporting Information Figure S34) the Ba and H bands are well-separated, which suggests a weak interaction between the hydrogen and barium sublattices. Electric conductivity is localized in H layers consist of quasi one-dimensional …H–H–H… chains which are kind of interconnected (Figure 3, Table S1)

Analyzing the electronic structure of $BaH_{12}$, we can conclude that despite the low electronic density at the Fermi level, this compound may demonstrate superconductivity. Barium dodecahydride is the first known molecular superhydride with metallic conductivity embedded in layers and 1D chains of molecular hydrogen.

## 4. Superconductivity

On the basis of powder diffraction and thermodynamic calculations, we cannot unambiguously determine the structure of the H-sublattice in $BaH_{12}$, which is essential for understanding superconductivity. To clarify this, we measured the electric resistance $R$ of barium hydride samples using a well-known four-probe technique in the temperature range of 2–300 K. At pressures of 90–140 GPa, all four $BaH_x$ samples synthesized in electric DACs behave as typical metals with an almost linear decrease of $R(T)$. At low temperatures the resistance of the samples drops sharply, indicating a possible superconducting transition at about 5–7 K below 130 GPa (Supporting Information Figure S38). The obtained data together with the measured Raman spectra and optical microscopy exclude low-symmetry $BaH_{12}$ semiconducting structures, leaving for consideration only metallic and semimetallic modifications (Supporting Information Figures S36 and S37).

For a more detailed study of superconductivity, we assembled an electric cell DAC #E5 with an 80 μm culet, $c$-BN/epoxy insulating gasket, 45×32 μm Ba piece, and sputtered 0.5 μm thick Mo electrodes. After the laser heating at 1600 K and 140 GPa, the Ba/AB sample demonstrated the superconducting transition at around 20 K (Figure 9a). When we tried to change the pressure, the cell collapsed and pressure dropped to 65 GPa.

Our calculations (Figure 9b) demonstrate that the low density of electronic states near the Fermi level in $Cmc2_1$-$BaH_{12}$ is associated with a weak electron-phonon coupling resulting in low λ = 0.67, $T_C$ = 19–32 K, and $μ_0H_{c2}(0)$ = 2.5–4.2 T at 150 GPa ($μ^*$ = 0.15–0.1). Dependence of $T_C$ on pressure has a slope $dT_C/dP$ = 0.6 K/GPa, and at 140 GPa expected $T_C$ is ~26 K, close to the experiment. Using the standard BCS formalism (Supporting Information), we also estimated the coherence length $ξ_{BCS}$ = 88 Å at 150 GPa ($μ^*$ = 0.1), superconducting gap of ~5 meV, and isotope coefficient β = 0.45.

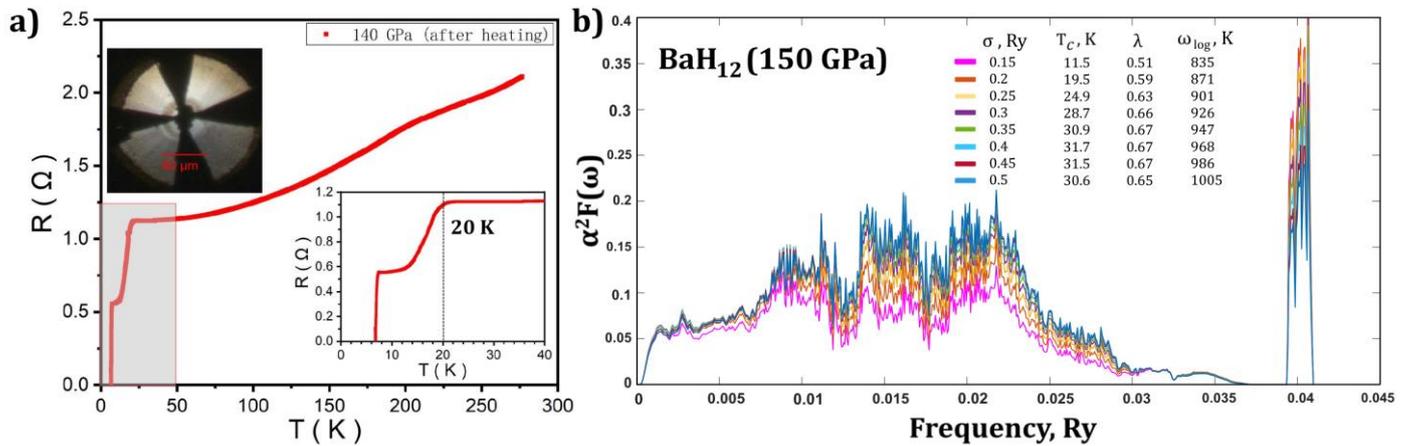

**Figure 9.** (a) Four-probe measurements of resistance $R(T)$ of the BaH$_x$ sample synthesized at 140 GPa. The superconducting transition was detected at ~20 K. (b) Calculated Eliashberg functions and electron-phonon coupling parameters for pseudocubic BaH$_{12}$ at 150 GPa (Goedecker–Hartwigsen–Hutter–Teter pseudopotential, $\mu^* = 0.1$).

As in all metal hydrides and polyhydrides, metal atoms donate some electrons to the hydrogen atoms. These electrons occupy the antibonding orbitals in the H$_2$ molecules and weaken the H-H bonds. If zero electrons are transferred, the H$_2$ molecules will persist and they will not contribute to superconductivity. If 1 electron is transferred, the hydride H$^-$ ions will be formed, with no H-H bonds and, again, little or no contribution to superconductivity. At intermediate electron doping levels (as it has been found [19, 21], the optimum is ~0.3 electrons) weak H-H bonds (e.g., as in clathrate polyhydrides) are formed.

In BaH$_{12}$, each H atom accepts few electrons, on average 0.16 electrons. As a result, H$_2$ and H$_3$ groups are still present in the structure, and we have a rather low $T_C$. We think that at high pressures, due to dissociation of molecular groups, BaH$_{12}$ may have network of weak H-H bonds (rather than discrete H$_2$ and H$_3$-groups) and, as a result, a much higher $T_C$. Increasing the pressure will also facilitate further metallization of BaH$_{12}$ and symmetrization of the hydrogen sublattice, increasing $N(E_F)$. To estimate the possible improvement, we calculated the superconducting parameters of $I4/mmm$-BaH$_{12}$ and $Fm\bar{3}m$-BaH$_{12}$, isostructural to YB$_{12}$, structures which were considered as possible solutions at the first step of the XRD interpretation. The calculations show that filling of pseudogap in $N(E)$ makes it possible to reach $T_C \sim 200$ K with $\lambda = 3$-$4.5$ in these compounds (see Supporting Figures S39-40).

## Conclusions

In our study of the high-pressure chemistry of the Ba–H system, we successfully synthesized in four independent DACs novel barium superhydride BaH$_{12}$ with a pseudocubic crystal structure, stabilized in the pressure range of 75–173 GPa. The compound was obtained by the laser heating of metallic barium with an excess of ammonia borane compressed to 173, 160, 146, and 90 GPa. The Ba sublattice structure of BaH$_{12}$ was resolved using the X-ray synchrotron diffraction, evolutionary structure prediction, and several postprocessing Python scripts, including an XRD matching algorithm. The discovered BaH$_{12}$ has the highest hydrogen content (>92 mol %) among all metal hydrides synthesized so far and unique metallic conductivity, localized in layers of molecular hydrogen. The experimentally established lower limit of stability of barium dodecahydride is 75 GPa. The third-order Birch–Murnaghan equation of state and unit cell parameters of BaH$_{12}$ were found in the pressure range of 75–173 GPa: $V_{100} = 45.47 \pm 0.13$ Å$^3$, $K_{100} = 305 \pm 8.5$ GPa, and $K'_{100} = 3.8 \pm 0.48$. The ab initio calculations confirm a small distortion of the ideal $fcc$-barium sublattice to $Cmc2_1$ (or even $P2_1$) space group, determined by the presence of additional weak reflections in the diffraction patterns. The impurity phase analysis indicates the possible presence of BaH$_6$ and BaH$_{10}$. According to the theoretical calculations and experimental measurements, BaH$_{12}$ exhibits semimetallic and superconducting properties with $T_C = 20$ K at 140 GPa, and its crystal structure contains H$_2$ and H$_3^-$ groups. The results of these experiments confirm that the comparative stability of superhydrides increases with the growth of the period number of a hydride-forming element in the periodic table.[21]

**Acknowledgments**

The authors thank the staff of the Shanghai Synchrotron Radiation Facility and express their gratitude to Bingbing Liu's group (Jilin University) for their help in the laser heating of samples. This work was supported by the National Key R&D Program of China (Grant No. 2018YFA0305900), the National Natural Science Foundation of China (Grant Nos. 11974133, 51720105007, 51632002, 11674122, 11574112, 11474127, and 11634004), the National Key Research and Development Program of China (Grant No. 2016YFB0201204), the Program for Changjiang Scholars and Innovative Research Team in University (Grant No. IRT_15R23), and the National Fund for Fostering Talents of Basic Science (Grant No. J1103202). A.R.O. thanks the Russian Science Foundation (Grant No. 19-72-30043). Portions of this work were performed at GeoSoilEnviroCARS (The University of Chicago, Sector 13), Advanced Photon Source (APS), Argonne National Laboratory. GeoSoilEnviroCARS is supported by the National Science Foundation – Earth Sciences (EAR – 1634415) and Department of Energy- GeoSciences (DE-FG02-94ER14466). This research used resources of the Advanced Photon Source, a U.S. Department of Energy (DOE) Office of Science User Facility operated for the DOE Office of Science by Argonne National Laboratory under Contract No. DE-AC02-06CH11357. A.G.K., D.V.S. thank the Russian Foundation for Basic Research (Grant No. 19-03-00100). A.F.G. acknowledges the support of the Army Research Office. We thank Dr. Christian Tantardini (Skoltech) for help with the calculation of the Bader charges, and Ms. Di Zhou (JLU) for preparing the convex hulls and $\alpha^2 F(\omega)$ diagrams.